\input epsf
\input amssym
\input eplain
\input\jobname.intref

\newfam\scrfam
\batchmode\font\tenscr=rsfs10 \errorstopmode
\ifx\tenscr\nullfont
        \message{rsfs script font not available. Replacing with calligraphic.}
        
\else   
        \font\sevenscr=rsfs7
        \font\fivescr=rsfs5
        \skewchar\tenscr='177 \skewchar\sevenscr='177 \skewchar\fivescr='177
        \textfont\scrfam=\tenscr \scriptfont\scrfam=\sevenscr
        \scriptscriptfont\scrfam=\fivescr

\fi
\catcode`\@=11
\newfam\frakfam
\batchmode\font\tenfrak=eufm10 \errorstopmode
\ifx\tenfrak\nullfont
        \message{eufm font not available. Replacing with italic.}
        \def\frak{\it}
\else
	
	\font\sevenfrak=eufm7 \font\fivefrak=eufm5
        
	\textfont\frakfam=\tenfrak
	\scriptfont\frakfam=\sevenfrak \scriptscriptfont\frakfam=\fivefrak
	\def\frak{\fam\frakfam}
\fi
\catcode`\@=\active
\newfam\msbfam
\batchmode\font\twelvemsb=msbm10 scaled\magstep1 \errorstopmode
\ifx\twelvemsb\nullfont\def\Bbb{\bf}
        
	\font\eightbbb=cmb10 at 8pt
	\message{Blackboard bold not available. Replacing with boldface.}
\else   \catcode`\@=11
        \font\tenmsb=msbm10 \font\sevenmsb=msbm7 \font\fivemsb=msbm5
        \textfont\msbfam=\tenmsb
        \scriptfont\msbfam=\sevenmsb \scriptscriptfont\msbfam=\fivemsb
        \def\Bbb{\relax\expandafter\Bbb@}
        \def\Bbb@#1{{\Bbb@@{#1}}}
        \def\Bbb@@#1{\fam\msbfam\relax#1}
        \catcode`\@=\active
	
	\font\eightbbb=msbm8
\fi
        \font\fivemi=cmmi5
        \font\sixmi=cmmi6
        \font\eightrm=cmr8              \def\xrm{\eightrm}
        \font\eightbf=cmbx8             \def\xbf{\eightbf}
        \font\eightit=cmti10 at 8pt     \def\xit{\eightit}

        \font\eighttt=cmtt8

        \font\eightcp=cmcsc8    
                      \def\xold{\eighti}
        \font\eightmi=cmmi8
                     \def\xbold{\eightib}
        \font\teni=cmmi10               \def\old{\teni}
        \font\tencp=cmcsc10

        \font\twelvecp=cmcsc10 scaled\magstep1
        
        \font\sixrm=cmr6
        \font\fiverm=cmr5

        \font\eightsy=cmsy8
        \font\sixsy=cmsy6
        \font\eightsl=cmsl8

        \font\sixbf=cmbx6

	 at10pt	
	\font\twelvehelvbold=phvb at12pt
	 at14pt
	\font\sixteenhelvbold=phvb at16pt
	 at16pt



\def\xbold{\xbf}
\def\xold{\xrm}


\def\noblackbox{\overfullrule=0pt}
\noblackbox

\def\eightpoint{
\def\rm{\fam0\eightrm}
\textfont0=\eightrm \scriptfont0=\sixrm \scriptscriptfont0=\fiverm
\textfont1=\eightmi  \scriptfont1=\sixmi  \scriptscriptfont1=\fivemi
\textfont2=\eightsy \scriptfont2=\sixsy \scriptscriptfont2=\fivesy
\textfont3=\tenex   \scriptfont3=\tenex \scriptscriptfont3=\tenex
\textfont\itfam=\eightit \def\it{\fam\itfam\eightit}
\textfont\slfam=\eightsl \def\sl{\fam\slfam\eightsl}
\textfont\ttfam=\eighttt \def\tt{\fam\ttfam\eighttt}
\textfont\bffam=\eightbf \scriptfont\bffam=\sixbf 
                         \scriptscriptfont\bffam=\fivebf
                         \def\bf{\fam\bffam\eightbf}
\normalbaselineskip=10pt}



\newtoks\headtext
\headline={\ifnum\pageno=1\hfill\else
	\ifodd\pageno{\eightcp\the\headtext}{ }\dotfill{ }{\old\folio}
	\else{\old\folio}{ }\dotfill{ }{\eightcp\the\headtext}\fi
	\fi}
\def\makeheadline{\vbox to 0pt{\vss\noindent\the\headline\break
\hbox to\hsize{\hfill}}
        \vskip2\baselineskip}
\newcount\infootnote
\infootnote=0
\newcount\footnotecount
\footnotecount=1
\def\foot#1{\infootnote=1
\footnote{${}^{\the\footnotecount}$}{\vtop{\baselineskip=.75\baselineskip
\advance\hsize by
-\parindent{\eightpoint\rm\hskip-\parindent
#1}\hfill\vskip\parskip}}\infootnote=0\global\advance\footnotecount by
1}
\newcount\refcount
\refcount=1
\newwrite\refwrite
\def\oldsize{\ifnum\infootnote=1\xold\else\old\fi}
\def\ref#1#2{
	\def#1{{{\oldsize\the\refcount}}\ifnum\the\refcount=1\immediate\openout\refwrite=\jobname.refs\fi\immediate\write\refwrite{\item{[{\xold\the\refcount}]} 
	#2\hfill\par\vskip-2pt}\xdef#1{{\noexpand\oldsize\the\refcount}}\global\advance\refcount by 1}
	}
\def\refout{\eightpoint\catcode`\@=11
        \xrm\immediate\closeout\refwrite
        \vskip2\baselineskip
        {\noindent\twelvecp References}\hfill\vskip\baselineskip
        \baselineskip=.75\baselineskip
        \input\jobname.refs
        \baselineskip=4\baselineskip \divide\baselineskip by 3
        \catcode`\@=\active\rm}

\def\skipref#1{\hbox to15pt{\phantom{#1}\hfill}\hskip-15pt}

\def\hepth#1{\href{http://xxx.lanl.gov/abs/hep-th/#1}{arXiv:\allowbreak
hep-th\slash{\xold#1}}}

\def\arxiv#1#2{\href{http://arxiv.org/abs/#1.#2}{arXiv:\allowbreak
{\xold#1}.{\xold#2}}} 
 
\def\jhep#1#2#3#4{\href{http://jhep.sissa.it/stdsearch?paper=#2\%28#3\%29#4}{J. High Energy Phys. {\xbold #1#2} ({\xold#3}) {\xold#4}}}

\def\CQG#1#2#3{Class. Quantum Grav. {\xbold#1} ({\xold#2}) {\xold#3}}
\def\FP#1#2#3{Fortsch. Phys. {\xbold#1} ({\xold#2}) {\xold#3}}

\def\JHEP{\jhep}

\def\JPA#1#2#3{J. Phys. {\xbf A}{\xbold#1} ({\xold#2}) {\xold#3}}

\def\MPLA#1#2#3{Mod. Phys. Lett. {\xbf A}{\xbold#1} ({\xold#2}) {\xold#3}}

\def\NPB#1#2#3{Nucl. Phys. {\xbf B}{\xbold#1} ({\xold#2}) {\xold#3}}

\def\PLB#1#2#3{Phys. Lett. {\xbf B}{\xbold#1} ({\xold#2}) {\xold#3}}

\def\PRD#1#2#3{Phys. Rev. {\xbf D}{\xbold#1} ({\xold#2}) {\xold#3}}
\def\PRL#1#2#3{Phys. Rev. Lett. {\xbold#1} ({\xold#2}) {\xold#3}}

\newcount\sectioncount
\sectioncount=0
\def\section#1#2{\global\eqcount=0
	\global\subsectioncount=0
        \global\advance\sectioncount by 1
	\ifnum\sectioncount>1
	        \vskip2\baselineskip
	\fi
\noindent{\twelvecp\the\sectioncount. #2}\par\nobreak
       \vskip.5\baselineskip\noindent
        \xdef#1{{\old\the\sectioncount}}}
\newcount\subsectioncount
\def\subsection#1#2{\global\advance\subsectioncount by 1
\vskip.75\baselineskip\noindent\line{\tencp\the\sectioncount.\the\subsectioncount. #2\hfill}\nobreak\vskip.4\baselineskip\nobreak\noindent\xdef#1{{\old\the\sectioncount}.{\old\the\subsectioncount}}}
\def\immediatesubsection#1#2{\global\advance\subsectioncount by 1
\vskip-\baselineskip\noindent
\line{\tencp\the\sectioncount.\the\subsectioncount. #2\hfill}
	\vskip.5\baselineskip\noindent
	\xdef#1{{\old\the\sectioncount}.{\old\the\subsectioncount}}}
\newcount\subsubsectioncount
\def\subsubsection#1#2{\global\advance\subsubsectioncount by 1
\vskip.75\baselineskip\noindent\line{\tencp\the\sectioncount.\the\subsectioncount.\the\subsubsectioncount. #2\hfill}\nobreak\vskip.4\baselineskip\nobreak\noindent\xdef#1{{\old\the\sectioncount}.{\old\the\subsectioncount}.{\old\the\subsubsectioncount}}}
\newcount\appendixcount
\appendixcount=0
\def\appendix#1{\global\eqcount=0
        \global\advance\appendixcount by 1
        \vskip2\baselineskip\noindent
        \ifnum\the\appendixcount=1
        {\twelvecp Appendix A: #1}\par\nobreak
                        \vskip.5\baselineskip\noindent\fi
        \ifnum\the\appendixcount=2
        {\twelvecp Appendix B: #1}\par\nobreak
                        \vskip.5\baselineskip\noindent\fi
        \ifnum\the\appendixcount=3
        {\twelvecp Appendix C: #1}\par\nobreak
                        \vskip.5\baselineskip\noindent\fi}
\def\acknowledgements{\immediate\write\contentswrite{\item{}\hbox
        to\contentlength{Acknowledgements\dotfill\the\pageno}}
        \vskip2\baselineskip\noindent
        \underbar{\it Acknowledgements:}\ }
\newcount\eqcount
\eqcount=0
\def\Eqn#1{\global\advance\eqcount by 1
\ifnum\the\sectioncount=0
	\xdef#1{{\noexpand\oldsize\the\eqcount}}
	\eqno({\oldstyle\the\eqcount})
\else
        \ifnum\the\appendixcount=0
\xdef#1{{\noexpand\oldsize\the\sectioncount}.{\noexpand\oldsize\the\eqcount}}
                \eqno({\oldstyle\the\sectioncount}.{\oldstyle\the\eqcount})\fi
        \ifnum\the\appendixcount=1
	        \xdef#1{{\noexpand\oldstyle A}.{\noexpand\oldstyle\the\eqcount}}
                \eqno({\oldstyle A}.{\oldstyle\the\eqcount})\fi
        \ifnum\the\appendixcount=2
	        \xdef#1{{\noexpand\oldstyle B}.{\noexpand\oldstyle\the\eqcount}}
                \eqno({\oldstyle B}.{\oldstyle\the\eqcount})\fi
        \ifnum\the\appendixcount=3
	        \xdef#1{{\noexpand\oldstyle C}.{\noexpand\oldstyle\the\eqcount}}
                \eqno({\oldstyle C}.{\oldstyle\the\eqcount})\fi
\fi}
\def\eqn{\global\advance\eqcount by 1
\ifnum\the\sectioncount=0
	\eqno({\oldstyle\the\eqcount})
\else
        \ifnum\the\appendixcount=0
                \eqno({\oldstyle\the\sectioncount}.{\oldstyle\the\eqcount})\fi
        \ifnum\the\appendixcount=1
                \eqno({\oldstyle A}.{\oldstyle\the\eqcount})\fi
        \ifnum\the\appendixcount=2
                \eqno({\oldstyle B}.{\oldstyle\the\eqcount})\fi
        \ifnum\the\appendixcount=3
                \eqno({\oldstyle C}.{\oldstyle\the\eqcount})\fi
\fi}
\def\multi{\global\advance\eqcount by 1}
\def\multieqn#1{({\oldstyle\the\sectioncount}.{\oldstyle\the\eqcount}\hbox{#1})}
\def\multiEqn#1#2{\xdef#1{{\oldstyle\the\sectioncount}.{\old\the\eqcount}#2}
        ({\oldstyle\the\sectioncount}.{\oldstyle\the\eqcount}\hbox{#2})}
\def\multiEqnAll#1{\xdef#1{{\oldstyle\the\sectioncount}.{\old\the\eqcount}}}
\newcount\tablecount
\tablecount=0
\def\Table#1#2#3{\global\advance\tablecount by 1
\immediate\write\intrefwrite{\def\noexpand#1{{\noexpand\oldsize\the\tablecount}}}
       \vtop{\vskip2\parskip
       \centerline{#2}
       \vskip5\parskip
       \centerline{\it Table \the\tablecount: #3}
       \vskip2\parskip}}
\newcount\figurecount
\figurecount=0
\def\Figure#1#2#3{\global\advance\figurecount by 1
\immediate\write\intrefwrite{\def\noexpand#1{{\noexpand\oldsize\the\figurecount}}}
       \vtop{\vskip2\parskip
       \centerline{#2}
       \vskip4\parskip
       \centerline{\it Figure \the\figurecount: #3}
       \vskip3\parskip}}
\newtoks\url
\def\Href#1#2{\catcode`\#=12\url={#1}\catcode`\#=\active#2}
\def\href#1#2{{#2}}

\parskip=3.5pt plus .3pt minus .3pt
\baselineskip=14pt plus .1pt minus .05pt
\lineskip=.5pt plus .05pt minus .05pt
\lineskiplimit=.5pt
\abovedisplayskip=18pt plus 4pt minus 2pt
\belowdisplayskip=\abovedisplayskip
\hsize=14cm
\vsize=20cm
\hoffset=1.5cm
\voffset=1.8cm
\frenchspacing
\footline={}
\raggedbottom

\newskip\origparindent
\origparindent=\parindent

\def\ss{\scriptstyle}

\def\*{\partial}
\def\punkt{\,\,.}
\def\komma{\,\,,}

\def\={\!=\!}
\def\small#1{{\hbox{$#1$}}}

\def\fraction#1{\small{1\over#1}}
\def\fr{\fraction}
\def\Fraction#1#2{\small{#1\over#2}}
\def\Fr{\Fraction}

\def\eg{{\it e.g.}}

\def\ie{{\it i.e.}}

\def\a{\alpha}

\def\Tr{\hbox{Tr}\,}

\def\dslash{\*\hskip-5.5pt/\hskip.5pt}


\def\appendix#1#2{\global\eqcount=0
        \global\advance\appendixcount by 1
        \vskip2\baselineskip\noindent
        \ifnum\the\appendixcount=1
        \immediate\write\intrefwrite{\def\noexpand#1{A}}
        {\twelvecp Appendix A: #2}\par\nobreak
                        \vskip.5\baselineskip\noindent\fi
        \ifnum\the\appendixcount=2
        {\twelvecp Appendix B: #2}\par\nobreak
                        \vskip.5\baselineskip\noindent\fi
        \ifnum\the\appendixcount=3
        {\twelvecp Appendix C: #2}\par\nobreak
                        \vskip.5\baselineskip\noindent\fi}

\def\lb{\bar\l}

\def\textfrac#1#2{\raise .45ex\hbox{\the\scriptfont0 #1}\nobreak\hskip-1pt/\hskip-1pt\hbox{\the\scriptfont0 #2}}


\def\frac{\Fr}

\def\mathbb{\Bbb}






\catcode`@=11
\def\openupnormal{\afterassignment\@penupnormal\dimen@=}
\def\@penupnormal{\advance\normallineskip\dimen@
  \advance\normalbaselineskip\dimen@
  \advance\normallineskiplimit\dimen@}
\catcode`@=12

\def\EqMatrix{\let\quad\enspace\openupnormal6pt\matrix}




\def\lb{\bar\l}

\def\textfrac#1#2{\raise .45ex\hbox{\the\scriptfont0 #1}\nobreak\hskip-1pt/\hskip-1pt\hbox{\the\scriptfont0 #2}}


\def\frac{\Fr}

\def\mathbb{\Bbb}

\newskip\scrskip
\scrskip=-.6pt plus 0pt minus .1pt


\newwrite\intrefwrite
\immediate\openout\intrefwrite=\jobname.intref

\newwrite\contentswrite

\newdimen\sublength
\sublength=\hsize 
\advance\sublength by -\parindent

\newdimen\contentlength
\contentlength=\sublength

\advance\sublength by -\parindent

\def\section#1#2{\global\eqcount=0
	\global\subsectioncount=0
        \global\advance\sectioncount by 1
\ifnum\the\sectioncount=1\immediate\openout\contentswrite=\jobname.contents\fi
	\ifnum\sectioncount>1
	        \vskip2\baselineskip
	\fi
\immediate\write\intrefwrite{\def\noexpand#1{{\noexpand\oldsize\the\sectioncount}}}
\immediate\write\contentswrite{\item{\the\sectioncount.}\hbox to\contentlength{#2\dotfill\the\pageno}}
\noindent{\twelvecp\the\sectioncount. #2}\par\nobreak
       \vskip.5\baselineskip\noindent}

\def\subsection#1#2{\global\advance\subsectioncount by 1
\immediate\write\contentswrite{\itemitem{\the\sectioncount.\the\subsectioncount.}\hbox
to\sublength{#2\dotfill\the\pageno}}
\immediate\write\intrefwrite{\def\noexpand#1{{\noexpand\oldsize\the\sectioncount}.{\noexpand\oldsize\the\subsectioncount}}}\vskip.75\baselineskip\noindent\line{\tencp\the\sectioncount.\the\subsectioncount. #2\hfill}\nobreak\vskip.4\baselineskip\nobreak\noindent}

\def\immediatesubsection#1#2{\global\advance\subsectioncount by 1
\immediate\write\contentswrite{\itemitem{\the\sectioncount.\the\subsectioncount.}\hbox
to\sublength{#2\dotfill\the\pageno}}
\immediate\write\intrefwrite{\def\noexpand#1{{\noexpand\oldsize\the\sectioncount}.{\noexpand\oldsize\the\subsectioncount}}}
\vskip-\baselineskip\noindent
\line{\tencp\the\sectioncount.\the\subsectioncount. #2\hfill}
	\vskip.5\baselineskip\noindent}

\def\contentsout{\catcode`\@=11
        \vskip2\baselineskip
        {\noindent\twelvecp Contents}\hfill\vskip\baselineskip
        \input\jobname.contents
        \catcode`\@=\active\rm
\vskip3\baselineskip
}

\def\refout{\eightpoint\catcode`\@=11
        \immediate\write\contentswrite{\item{}\hbox to\contentlength{References\dotfill\the\pageno}}
        \xrm\immediate\closeout\refwrite
        \vskip2\baselineskip
        {\noindent\twelvecp References}\hfill\vskip\baselineskip
        \baselineskip=.75\baselineskip
        \input\jobname.refs
        \baselineskip=4\baselineskip \divide\baselineskip by 3
        \catcode`\@=\active\rm}


\ref\HatsudaKamimuraSiegelI{M. Hatsuda, K. Kamimura and W. Siegel,
{\xit ``Superspace with manifest T-duality from type II
superstring''}, \jhep{14}{06}{2014}{039} [\arxiv{1403}{3887}].}

\ref\HatsudaKamimuraSiegelII{M. Hatsuda, K. Kamimura and W. Siegel,
{\xit ``Ramond-Ramond gauge fields in superspace with manifest
T-duality''}, \jhep{15}{02}{2015}{134} [\arxiv{1411}{2206}].} 

\ref\Duff{M.J. Duff, {\xit ``Duality rotations in string
theory''}, \NPB{335}{1990}{610}.}

\ref\Tseytlin{A.A.~Tseytlin,
  {\xit ``Duality symmetric closed string theory and interacting
  chiral scalars''}, 
  \NPB{350}{1991}{395}.}

\ref\SiegelI{W.~Siegel,
  {\xit ``Two vierbein formalism for string inspired axionic gravity''},
  \PRD{47}{1993}{5453}
  [\hepth{9302036}].}

\ref\SiegelII{ W.~Siegel,
  {\xit ``Superspace duality in low-energy superstrings''},
  \PRD{48}{1993}{2826}
  [\hepth{9305073}].}

\ref\SiegelIII{W.~Siegel,
  {\xit ``Manifest duality in low-energy superstrings''},
  in Berkeley 1993, Proceedings, Strings '93 353
  [\hepth{9308133}].}

\ref\HullDoubled{C.M. Hull, {\xit ``Doubled geometry and
T-folds''}, \jhep{07}{07}{2007}{080}
[\hepth{0605149}].}

\ref\HullT{C.M. Hull, {\xit ``A geometry for non-geometric string
backgrounds''}, \jhep{05}{10}{2005}{065} [\hepth{0406102}].}

\ref\HullM{C.M. Hull, {\xit ``Generalised geometry for M-theory''},
\jhep{07}{07}{2007}{079} [\hepth{0701203}].}

\ref\HullZwiebachDFT{C. Hull and B. Zwiebach, {\xit ``Double field
theory''}, \jhep{09}{09}{2009}{99} [\arxiv{0904}{4664}].}

\ref\HohmHullZwiebachI{O. Hohm, C.M. Hull and B. Zwiebach, {\xit ``Background
independent action for double field
theory''}, \jhep{10}{07}{2010}{016} [\arxiv{1003}{5027}].}

\ref\HohmHullZwiebachII{O. Hohm, C.M. Hull and B. Zwiebach, {\xit
``Generalized metric formulation of double field theory''},
\jhep{10}{08}{2010}{008} [\arxiv{1006}{4823}].} 

\ref\HohmKwak{O. Hohm and S.K. Kwak, {\xit ``$N=1$ supersymmetric
double field theory''}, \jhep{12}{03}{2012}{080} [\arxiv{1111}{7293}].}

\ref\HohmKwakFrame{O. Hohm and S.K. Kwak, {\xit ``Frame-like geometry
of double field theory''}, \JPA{44}{2011}{085404} [\arxiv{1011}{4101}].}

\ref\JeonLeeParkI{I. Jeon, K. Lee and J.-H. Park, {\xit ``Differential
geometry with a projection: Application to double field theory''},
\jhep{11}{04}{2011}{014} [\arxiv{1011}{1324}].}

\ref\JeonLeeParkII{I. Jeon, K. Lee and J.-H. Park, {\xit ``Stringy
differential geometry, beyond Riemann''}, 
\PRD{84}{2011}{044022} [\arxiv{1105}{6294}].}

\ref\JeonLeeParkIII{I. Jeon, K. Lee and J.-H. Park, {\xit
``Supersymmetric double field theory: stringy reformulation of supergravity''},
\PRD{85}{2012}{081501} [\arxiv{1112}{0069}].}

\ref\HohmZwiebachLarge{O. Hohm and B. Zwiebach, {\xit ``Large gauge
transformations in double field theory''}, \jhep{13}{02}{2013}{075}
[\arxiv{1207}{4198}].} 

\ref\Park{J.-H.~Park,
  {\xit ``Comments on double field theory and diffeomorphisms''},
  \jhep{13}{06}{2013}{098}
  [\arxiv{1304}{5946}].}

\ref\BermanCederwallPerry{D.S. Berman, M. Cederwall and M.J. Perry,
{\xit ``Global aspects of double geometry''}, 
\jhep{14}{09}{2014}{66} [\arxiv{1401}{1311}].}

\ref\CederwallGeometryBehind{M. Cederwall, {\xit ``The geometry behind
double geometry''}, 
\jhep{14}{09}{2014}{70} [\arxiv{1402}{2513}].}

\ref\PachecoWaldram{P.P. Pacheco and D. Waldram, {\xit ``M-theory,
exceptional generalised geometry and superpotentials''},
\jhep{08}{09}{2008}{123} [\arxiv{0804}{1362}].}

\ref\Hillmann{C. Hillmann, {\xit ``Generalized $E_{7(7)}$ coset
dynamics and $D=11$ supergravity''}, \jhep{09}{03}{2009}{135}
[\arxiv{0901}{1581}].}

\ref\BermanPerryGen{D.S. Berman and M.J. Perry, {\xit ``Generalised
geometry and M-theory''}, \jhep{11}{06}{2011}{074} [\arxiv{1008}{1763}].}    

\ref\BermanGodazgarPerry{D.S. Berman, H. Godazgar and M.J. Perry,
{\xit ``SO(5,5) duality in M-theory and generalized geometry''},
\PLB{700}{2011}{65} [\arxiv{1103}{5733}].} 

\ref\BermanMusaevPerry{D.S. Berman, E.T. Musaev and M.J. Perry,
{\xit ``Boundary terms in generalized geometry and doubled field theory''},
\PLB{706}{2011}{228} [\arxiv{1110}{97}].} 

\ref\BermanGodazgarGodazgarPerry{D.S. Berman, H. Godazgar, M. Godazgar  
and M.J. Perry,
{\xit ``The local symmetries of M-theory and their formulation in
generalised geometry''}, \jhep{12}{01}{2012}{012}
[\arxiv{1110}{3930}].} 

\ref\BermanGodazgarPerryWest{D.S. Berman, H. Godazgar, M.J. Perry and
P. West,
{\xit ``Duality invariant actions and generalised geometry''}, 
\jhep{12}{02}{2012}{108} [\arxiv{1111}{0459}].} 

\ref\CoimbraStricklandWaldram{A. Coimbra, C. Strickland-Constable and
D. Waldram, {\xit ``$E_{d(d)}\times\hbox{\eightbbb R}^+$ generalised geometry,
connections and M theory'' }, \jhep{14}{02}{2014}{054} [\arxiv{1112}{3989}].} 

\ref\CoimbraStricklandWaldramII{A. Coimbra, C. Strickland-Constable and
D. Waldram, {\xit ``Supergravity as generalised geometry II:
$E_{d(d)}\times\hbox{\eightbbb R}^+$ and M theory''}, 
\jhep{14}{03}{2014}{019} [\arxiv{1212}{1586}].}  

\ref\JeonLeeParkSuh{I. Jeon, K. Lee, J.-H. Park and Y. Suh, {\xit
``Stringy unification of Type IIA and IIB supergravities under N=2
D=10 supersymmetric double field theory''}, \PLB{723}{2013}{245}
[\arxiv{1210}{5048}].} 

\ref\JeonLeeParkRR{I. Jeon, K. Lee and J.-H. Park, {\xit
``Ramond--Ramond cohomology and O(D,D) T-duality''},
\jhep{12}{09}{2012}{079} [\arxiv{1206}{3478}].} 

\ref\BermanCederwallKleinschmidtThompson{D.S. Berman, M. Cederwall,
A. Kleinschmidt and D.C. Thompson, {\xit ``The gauge structure of
generalised diffeomorphisms''}, \jhep{13}{01}{2013}{64} [\arxiv{1208}{5884}].}

\ref\ParkSuh{J.-H. Park and Y. Suh, {\xit ``U-geometry: SL(5)''},
\jhep{14}{06}{2014}{102} [\arxiv{1302}{1652}].} 

\ref\CederwallI{M.~Cederwall, J.~Edlund and A.~Karlsson,
  {\xit ``Exceptional geometry and tensor fields''},
  \jhep{13}{07}{2013}{028}
  [\arxiv{1302}{6736}].}

\ref\CederwallII{ M.~Cederwall,
  {\xit ``Non-gravitational exceptional supermultiplets''},
  \jhep{13}{07}{2013}{025}
  [\arxiv{1302}{6737}].}

\ref\SambtlebenHohmI{O.~Hohm and H.~Samtleben,
  {\xit ``Exceptional field theory I: $E_{6(6)}$ covariant form of
  M-theory and type IIB''}, 
  \PRD{89}{2014}{066016} [\arxiv{1312}{0614}].}

\ref\SambtlebenHohmII{O.~Hohm and H.~Samtleben,
  {\xit ``Exceptional field theory II: $E_{7(7)}$''},
  \PRD{89}{2014}{066016} [\arxiv{1312}{4542}].}

\ref\HohmSamtlebenIII{O. Hohm and H. Samtleben, {\xit ``Exceptional field
theory III: $E_{8(8)}$''}, \PRD{90}{2014}{066002} [\arxiv{1406}{3348}].}

\ref\KachruNew{S. Kachru, M.B. Schulz, P.K. Tripathy and S.P. Trivedi,
{\xit ``New supersymmetric string compactifications''}, 
\jhep{03}{03}{2003}{061} [\hepth{0211182}].}

\ref\Condeescu{C. Condeescu, I. Florakis, C. Kounnas and D. L\"ust, 
{\xit ``Gauged supergravities and non-geometric $Q$/$R$-fluxes from
asymmetric orbifold CFT's''}, 
\jhep{13}{10}{2013}{057} [\arxiv{1307}{0999}].}

\ref\CederwallUfoldBranes{M. Cederwall, {\xit ``M-branes on U-folds''},
in proceedings of 7th International Workshop ``Supersymmetries and
Quantum Symmetries'' Dubna, 2007 [\arxiv{0712}{4287}].}

\ref\HasslerLust{F. Hassler and D. L\"ust, {\xit ``Consistent
compactification of double field theory on non-geometric flux
backgrounds''}, \jhep{14}{05}{2014}{085} [\arxiv{1401}{5068}].}

\ref\CederwallDuality{M. Cederwall, {\xit ``T-duality and
non-geometric solutions from double geometry''}, \FP{62}{2014}{942}
[\arxiv{1409}{4463}].} 

\ref\CederwallRosabal{M. Cederwall and J.A. Rosabal, ``$E_8$
geometry'', \jhep{15}{07}{2015}{007}, [\arxiv{1504}{04843}].}

\ref\HohmKwakZwiebachI{O. Hohm, S.K. Kwak and B. Zwiebach, {\xit
``Unification of type II strings and T-duality''}, \PRL{107}{2011}{171603}
[\arxiv{1106}{5452}].}  

\ref\HohmKwakZwiebachII{O. Hohm, S.K. Kwak and B. Zwiebach, {\xit
``Double field theory of type II strings''}, \jhep{11}{09}{2011}{013}
[\arxiv{1107}{0008}].}  

\ref\HohmZwiebachGeometry{O. Hohm and B. Zwiebach, {\xit ``Towards an
invariant geometry of double field theory''}, \arxiv{1212}{1736}.} 

\ref\CGNT{M. Cederwall, U. Gran, B.E.W. Nilsson and D. Tsimpis,
{\xit ``Supersymmetric corrections to eleven-dimen\-sional supergravity''},
\jhep{05}{05}{2005}{052} [\hepth{0409107}].}

\ref\CartanSpinors{E. Cartan, {\xit ``Le\hskip.5pt,\hskip-3.5pt cons sur
la th\'eorie des spineurs''} (Hermann, Paris, 1937).}

\ref\AsakawaSasaWatamura{T. Asakawa, S. Sasa and S. Watamura, {\xit
``D-branes in generalized geometry and Dirac--Born--Infeld action''},
\jhep{12}{10}{2012}{064} [\arxiv{1206}{6964}].} 

\ref\BermanCederwallMalek{D.S. Berman, M. Cederwall and E. Malek,
{\xit work in progress}.}

\ref\DBranesII{M. Cederwall, A. von Gussich, B.E.W. Nilsson, P. Sundell
 and A. Westerberg,
{\xit ``The Dirichlet super-p-branes in ten-dimensional type IIA and IIB 
supergravity''},
\NPB{490}{1997}{179} [\hepth{9611159}].}

\ref\AzcarragaTownsend{J.A. de Azc\'arraga and P.K. Townsend, {\xit
``Superspace geometry and classification of supersymmetric extended
objects''}, \PRL{62}{1989}{2579}.} 

\ref\BerkovitsI{N. Berkovits, 
{\xit ``Super-Poincar\'e covariant quantization of the superstring''}, 
\jhep{00}{04}{2000}{018} [\hepth{0001035}].}

\ref\BerkovitsParticle{N. Berkovits, {\xit ``Covariant quantization of
the superparticle using pure spinors''}, \jhep{01}{09}{2001}{016}
[\hepth{0105050}].}

\ref\CederwallNilssonTsimpisI{M. Cederwall, B.E.W. Nilsson and D. Tsimpis,
{\xit ``The structure of maximally supersymmetric super-Yang--Mills
theory --- constraining higher order corrections''},
\jhep{01}{06}{2001}{034} 
[\hepth{0102009}].}

\ref\CederwallNilssonTsimpisII{M. Cederwall, B.E.W. Nilsson and D. Tsimpis,
{\xit ``D=10 super-Yang--Mills at $\ss O(\a'^2)$''},
\JHEP{01}{07}{2001}{042} [\hepth{0104236}].}

\ref\SpinorialCohomology{M. Cederwall, B.E.W. Nilsson and D. Tsimpis, 
{\xit ``Spinorial cohomology and maximally supersymmetric theories''},
\jhep{02}{02}{2002}{009} [\hepth{0110069}];
M. Cederwall, {\xit ``Superspace methods in string theory,
supergravity and gauge theory''}, Lectures at the XXXVII Winter School
in Theoretical Physics ``New Developments in Fundamental Interactions
Theories'',  Karpacz, Poland,  Feb. 6-15, 2001, \hepth{0105176}.}

\ref\CederwallBLG{M. Cederwall, {\xit ``N=8 superfield formulation of
the Bagger--Lambert--Gustavsson model''}, \jhep{08}{09}{2008}{116}
[\arxiv{0808}{3242}].}

\ref\CederwallABJM{M. Cederwall, {\xit ``Superfield actions for N=8 
and N=6 conformal theories in three dimensions''},
\jhep{08}{10}{2008}{70}
[\arxiv{0809}{0318}].}

\ref\PureSGI{M. Cederwall, {\xit ``Towards a manifestly supersymmetric
    action for D=11 supergravity''}, \jhep{10}{01}{2010}{117}
    [\arxiv{0912}{1814}].}  

\ref\PureSGII{M. Cederwall, 
{\xit ``D=11 supergravity with manifest supersymmetry''},
    \MPLA{25}{2010}{3201} [\arxiv{1001}{0112}].}

\ref\PureSpinorOverview{M. Cederwall, {\xit ``Pure spinor superfields
--- an overview''}, Springer Proc. Phys. {\xbf153} ({\xrm2013}) {\xrm61} 
[\arxiv{1307}{1762}].}

\ref\KacSuperalgebras{V.G. Kac, {\xit ``Classification of simple Lie
superalgebras''}, Funktsional. Anal. i Prilozhen. {\xbold9}
({\xold1975}) {\xold91}.}

\ref\CederwallExceptionalTwistors{M. Cederwall, {\xit ``Twistors and
supertwistors for exceptional field
theory''}, \jhep{15}{12}{2015}{123} [\arxiv{1510}{02298}].}

\ref\MaDBrane{C.-T. Ma, {\xit ``Gauge transformation of double field
theory for open string''}, \PRD{92}{2015}{066004} [\arxiv{1411}{0287}].}

\ref\BandosStringsSuperspace{I Bandos, {\xit ``Strings in doubled
superspace''}, \PLB{751}{2015}{402} [\arxiv{1507}{07779}].}

\ref\BandosESeven{I Bandos, {\xit ``On section conditions of
$E_{7(+7)}$ exceptional field theory and superparticle in N=8 central
charge superspace''}, \jhep{16}{01}{2016}{132} [\arxiv{1512}{02287}].}


\ref\CederwallNilssonSix{M. Cederwall and B.E.W. Nilsson, {\xit ``Pure
spinors and D=6 super-Yang--Mills''}, \arxiv{0801}{1428}.}

\ref\CGNN{M. Cederwall, U. Gran, M. Nielsen and B.E.W. Nilsson, 
{\xit ``Manifestly supersymmetric M-theory''}, 
\JHEP{00}{10}{2000}{041} [\hepth{0007035}];
{\xit ``Generalised 11-dimensional supergravity''}, \hepth{0010042}.}

\ref\CederwallNilssonTsimpisI{M. Cederwall, B.E.W. Nilsson and D. Tsimpis,
{\xit ``The structure of maximally supersymmetric super-Yang--Mills
theory --- constraining higher order corrections''},
\jhep{01}{06}{2001}{034} 
[\hepth{0102009}].}

\ref\CederwallNilssonTsimpisII{M. Cederwall, B.E.W. Nilsson and D. Tsimpis,
{\xit ``D=10 super-Yang--Mills at $\ss O(\a'^2)$''},
\JHEP{01}{07}{2001}{042} [\hepth{0104236}].}

\ref\SpinorialCohomology{M. Cederwall, B.E.W. Nilsson and D. Tsimpis, 
{\xit ``Spinorial cohomology and maximally supersymmetric theories''},
\jhep{02}{02}{2002}{009} [\hepth{0110069}].}

\ref\SuperspaceMethods{M. Cederwall,
{\xit ``Superspace methods in string theory,
supergravity and gauge theory''}, Lectures at the XXXVII Winter School
in Theoretical Physics ``New Developments in Fundamental Interactions
Theories'',  Karpacz, Poland,  Feb. 6-15, 2001, \hepth{0105176}.}

\ref\CederwallNilssonTsimpisIII{M. Cederwall, B.E.W. Nilsson and D. Tsimpis,
{\xit ``Spinorial cohomology of abelian D=10 super-Yang--Mills at $\ss
O(\a'^3)$''}, 
\JHEP{02}{11}{2002}{023} [\hepth{0205165}].}

\ref\CGNT{M. Cederwall, U. Gran, B.E.W. Nilsson and D. Tsimpis,
{\xit ``Supersymmetric corrections to eleven-dimen\-sional supergravity''},
\jhep{05}{05}{2005}{052} [\hepth{0409107}].}

\ref\CederwallBLG{M. Cederwall, {\xit ``N=8 superfield formulation of
the Bagger--Lambert--Gustavsson model''}, \jhep{08}{09}{2008}{116}
[\arxiv{0808}{3242}].}

\ref\CederwallABJM{M. Cederwall, {\xit ``Superfield actions for N=8 
and N=6 conformal theories in three dimensions''},
\jhep{08}{10}{2008}{70}
[\arxiv{0809}{0318}].}

\ref\CederwallThreeConf{M. Cederwall, {\xit ``Pure spinor superfields,
with application to D=3 conformal models''}, 
Proc. Estonian Acad. Sci. {\xbf4} ({\xold2010})
[\arxiv{0906}{5490}].}

\ref\PureSGI{M. Cederwall, {\xit ``Towards a manifestly supersymmetric
    action for D=11 supergravity''}, \jhep{10}{01}{2010}{117}
    [\arxiv{0912}{1814}].}  

\ref\PureSGII{M. Cederwall, 
{\xit ``D=11 supergravity with manifest supersymmetry''},
    \MPLA{25}{2010}{3201} [\arxiv{1001}{0112}].}

\ref\SupergeometryPureSpinors{M. Cederwall, {\xit ``From supergeometry
to pure spinors''}, in the proceedings of the 6th Mathematical physics
meeting, Belgrade, September 2010, \arxiv{1012}{3334}.}

\ref\CederwallKarlssonBI{M. Cederwall and A. Karlsson, {\xit ``Pure
spinor superfields and Born--Infeld theory''},
\jhep{11}{11}{2011}{134} [\arxiv{1109}{0809}].}

\ref\CederwallPureSpinorSpace{M. Cederwall, {\xit ``The geometry of
pure spinor space''}, \jhep{12}{01}{2012}{150}  
\hfill\break[\arxiv{1111}{1932}].}

\ref\CederwallKarlssonLoop{M. Cederwall and A. Karlsson, {\xit ``Loop
amplitudes in maximal supergravity with manifest supersymmetry''},
\jhep{13}{03}{2013}{114} [\arxiv{1212}{5175}].}

\ref\PureSpinorOverview{M. Cederwall, {\xit ``Pure spinor superfields
--- an overview''}, Springer Proc. Phys. {\xbf153} ({\xrm2013}) {\xrm61} 
[\arxiv{1307}{1762}].}

\ref\ChangDeformationsI{C.-M. Chang, Y.-H. Lin, Y. Wang and X. Yin,
{\xit ``Deformations with maximal supersymmetries part 1: on-shell
formulation''},  \arxiv{1403}{0545}.}

\ref\ChangDeformationsII{C.-M. Chang, Y.-H. Lin, Y. Wang and X. Yin,
{\xit ``Deformations with maximal supersymmetries part 2: off-shell
formulation''}, \jhep{16}{04}{2016}{171} [\arxiv{1403}{0709}].}

\ref\BjornssonGreen{J. Bj\"ornsson and M.B. Green, {\xit ``5 loops in
25/4 dimensions''}, \jhep{10}{08}{2010}{132} [\arxiv{1004}{2692}].}

\ref\BjornssonMultiLoop{J. Bj\"ornsson, {\xit ``Multi-loop amplitudes
in maximally supersymmetric pure spinor field
theory''}, \jhep{11}{01}{2011}{002} [\arxiv{1009}{5906}].}

\ref\GrassiVanhoveAmpl{P.A. Grassi and P. Vanhove, {\xit ``Higher-loop
amplitudes in the non-minimal pure spinor
formalism''}, \jhep{09}{05}{2009}{089} [\arxiv{0903}{3903}].}

\ref\GIKOS{A. Galperin, E. Ivanov, S. Kalitzin, V. Ogievetsky and
E. Sokatchev, {\xit ``Unconstrained $N=2$ matter,
Yang--Mills and supergravity theories in harmonic
superspace''}, \CQG1{1984}{469}.} 

\ref\ChicherinSokatchevI{D. Chicherin and E. Sokatchev, {\xit ``$N=4$
super-Yang--Mills in LHC superspace. Part I: Classical and quantum
theory''}, \jhep{17}{02}{2017}{062} [\arxiv{1601}{06803}].}

\ref\ChicherinSokatchevII{D. Chicherin and E. Sokatchev, {\xit ``$N=4$
super-Yang--Mills in LHC superspace. Part II: Non-chiral correlation functions
of the stress-tensor multiplet''}, \jhep{17}{03}{2017}{048}
[\arxiv{1601}{06804}].} 

\ref\BerkovitsGreenRussoVanhove{N. Berkovits, M.B. Green, J. Russo and
    P. Vanhove, {\xit ``Non-renormalization conditions for four-gluon
    scattering in supersymmetric string and field
    theory''}, \jhep{09}{11}{2009}{063} [\arxiv{0908}{1923}].}

\ref\BerkovitsNonMinimal{N. Berkovits,
{\xit ``Pure spinor formalism as an N=2 topological string''},
\jhep{05}{10}{2005}{089} [\hepth{0509120}].}

\ref\BerkovitsNekrasovMultiloop{N. Berkovits and N. Nekrasov, {\xit
    ``Multiloop superstring amplitudes from non-minimal pure spinor
    formalism''}, \jhep{06}{12}{2006}{029} [\hepth{0609012}].}

\ref\ChangDeformationsII{C.-M. Chang, Y.-H. Lin, Y. Wang and X. Yin,
{\xit ``Deformations with maximal supersymmetries part 2: off-shell
formulation''}, \jhep{16}{04}{2016}{171} [\arxiv{1403}{0709}].}

\ref\CederwallPalmkvistBorcherds{M. Cederwall and J. Palmkvist, {\xit
``Superalgebras, constraints and partition functions''},
\jhep{08}{15}{2015}{36} [\arxiv{1503}{06215}].}


\def\lb{\bar\lambda}
\def\mb{\bar\mu}

%
\line{
\epsfysize=18mm
\epsffile{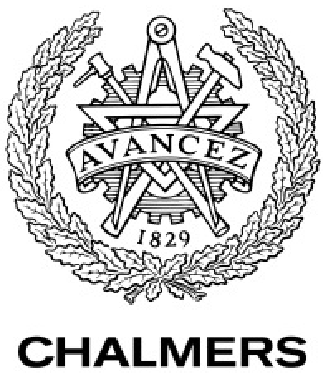}
\hfill}
\vskip-16mm

\line{\hfill}
\line{\hfill Gothenburg preprint}
\line{\hfill July, {\old2017}}
\line{\hrulefill}


\headtext={Cederwall: 
``An off-shell superspace reformulation...''}

\vfill

\centerline{\sixteenhelvbold
An off-shell superspace reformulation of}

\vskip3\parskip

\centerline{\sixteenhelvbold
D=4, N=4 super-Yang--Mills theory}

%

\vfill

\centerline{\twelvehelvbold Martin Cederwall}

\vfill
\vskip-1cm

\centerline{\it Division for Theoretical Physics}
\centerline{\it Department of Physics}
\centerline{\it Chalmers University of Technology}
\centerline{\it SE 412 96 Gothenburg, Sweden}

\vfill

{\narrower\noindent \underbar{Abstract:}
$D=4$, $N=4$ super-Yang--Mills theory has an off-shell superspace
formulation in terms of pure spinor superfields, which is directly
inherited from the $D=10$ theory. That superspace, in particular the
choice of pure spinor variables, is less suitable for dealing with
fields that are inherently 4-dimensional, such as the superfields
based on the scalars, which
are gauge-covariant, and traces of powers of scalars, which are
gauge-invariant. We give a reformulation of $D=4$, $N=4$ super-Yang--Mills
theory
in $N=4$ superspace, using
inherently 4-dimensional pure spinors. All
local degrees of freedom reside in a superfield based on the
physical scalars. The formalism should be suited for calculations of
correlators of traces of scalar superfields.

\smallskip}
\vfill

\font\xxtt=cmtt6

\vtop{\baselineskip=.6\baselineskip\xxtt
\line{\hrulefill}
\catcode`\@=11
\line{email: martin.cederwall@chalmers.se\hfill}
\catcode`\@=\active
}

\eject


\contentsout

\section\IntroductionSection{Introduction}Pure spinor superfields
provide the only known way to achieve off-shell superspace formulations
of maximally supersymmetric models.
The formalism originates in superstring theory [\BerkovitsI] and in
investigations of the ``ordinary'' superspace constraints
[\CederwallNilssonTsimpisI,\CederwallNilssonTsimpisII,\SpinorialCohomology].
$D=10$ super-Yang--Mills
theory (SYM) is the standard example
[\BerkovitsParticle,\CederwallKarlssonBI,\PureSpinorOverview],
but actions have been given for a number of other maximally
supersymmetric theories
[\PureSpinorOverview\skipref\CederwallBLG\skipref\CederwallABJM\skipref\CederwallThreeConf\skipref\PureSGI-\PureSGII], including $D=11$ supergravity.
The resulting field theories can be quantised, and yield Feynman rules
which lead to the most powerful convergence estimates for $D=4$, $N=4$
SYM and $D=4$, $N=8$ supergravity
[\GrassiVanhoveAmpl\skipref\BerkovitsGreenRussoVanhove\skipref\BjornssonGreen\skipref\BjornssonMultiLoop-\CederwallKarlssonLoop].

The focus of the present paper is $D=4$, $N=4$ SYM. Although the
dimensional reduction of $D=10$ SYM provides a good description of the
theory, it is less suited for some types of questions. This concerns
especially the treatment of gauge-invariant multiplets like the stress
tensor multiplet or the Konishi multiplet, which have no counterpart
in the $D=10$ theory. Such operators have been treated in harmonic
superspace [\GIKOS], especially Lorentz harmonic superspace
[\ChicherinSokatchevI,\ChicherinSokatchevII]. It would be desirable to
have access to a maximally supersymmetric superspace formulation which
is adapted to this kind of problem. The aim of this paper is to
provide a first step towards such a formalism in the form of a
classical action.

\vfill\eject

\section\TenSYMSection{Off-shell $D=10$ SYM and dimensional
reduction}The action for $D=10$ super-Yang--Mills theory is
$$
S=\int[dZ]\,\Tr\left(\fr2\Psi Q\Psi+\fr3\Psi^3\right)\punkt\Eqn\DTenAction
$$
For more detail than given below, see ref. [\PureSpinorOverview] and
references in that paper.
Here, $\Psi$ is
a scalar fermionic superfield $\Psi=\Psi(x,\theta,\lambda,\lb,d\lb)$,
depending on the non-minimal pure spinor
variables [\BerkovitsNonMinimal,\BerkovitsNekrasovMultiloop].
$\Psi$ carries ghost number 1.
The integration measure is
$$
[dZ]=d^{10}x\,d^{16}\theta\,\Omega\komma\eqn
$$
where $\Omega$ is the holomorphic 11-form on complex pure spinor
space, which is a non-compact Calabi--Yau space
[\CederwallPureSpinorSpace]. This integration needs regulation, see
Section \PureSpinorSection.
The BRST operator is
$$
Q=q+\bar\*=\lambda^\alpha
D_\alpha+d\lb_\alpha{\*\over\*\lb_\alpha}\punkt\eqn 
$$
The pure spinor $\lambda$ satisfies
$$
(\lambda\gamma^m\lambda)=0\punkt\eqn
$$
The second term in the BRST operator, the Dolbeault operator, does not
affect the cohomology [\BerkovitsNonMinimal], but is needed for a
non-degenerate integration measure, and for the construction of
operators with negative ghost number.

The zero-mode cohomology of $\Psi$ shows that it contains the SYM
fields, and in addition the ghost and antifields\foot{This and other
lists of cohomologies in 
the paper may be calculated by hand or by the computer-aided method of
ref. [\SpinorialCohomology].}. The proper
interpretation of the action (\DTenAction) is as a Batalin--Vilkovisky (BV)
action.
The cohomology is listed in Table \TenSYMTable, where Dynkin labels
with the conventions of Figure \DFiveDynkin\ are used.
When cohomologies are listed, the representations and quantum numbers
are those of the component fields.

\Figure\DFiveDynkin{\epsffile{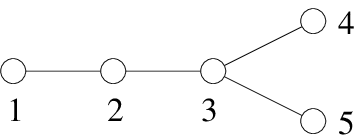}}{Dynkin diagram for $D_5$.}

\Table\TenSYMTable{$$
\vtop{\eightpoint\baselineskip16pt\lineskip0pt
\ialign{
$\hfill#\quad$&$\,\hfill#\hfill\,$&$\,\hfill#\hfill\,$&$\,\hfill#\hfill\,$
&$\,\hfill#\hfill\,$&$\,\hfill#\hfill$\cr
            \hbox{gh\#}=&1    &0    &-1    &-2   &-3\cr
\hbox{dim}=0&(00000)&\phantom{(00000)}&\phantom{(00000)}
&\phantom{(00000)}&\phantom{(00000)}\cr
        \fr2&\bullet&\bullet&       &       &    \cr
           1&\bullet&(10000)&\bullet&       &    \cr
       \Fr32&\bullet&(00001)&\bullet&\bullet&      \cr
           2&\bullet&\bullet&\bullet&\bullet&\bullet\cr
       \Fr52&\bullet&\bullet&(00010)&\bullet&\bullet\cr
           3&\bullet&\bullet&(10000)&\bullet&\bullet\cr
       \Fr72&\bullet&\bullet&\bullet&\bullet&\bullet\cr
           4&\bullet&\bullet&\bullet&(00000)&\bullet\cr
       \Fr92&\bullet&\bullet&\bullet&\bullet&\bullet\cr
}}
$$}{The zero-mode cohomology of $D=10$ SYM.}

Now, consider the dimensional reduction to $D=4$, $N=4$ SYM.
We split the 10-dimensional vector index as $m=(i,a)$, $i=0,\ldots,3$,
$a=1,\ldots,6$. Chiral spinors split as
$(00001)\rightarrow(0)(1)(001)\oplus(1)(0)(010)$, using the
conventions of Figure \DTwoDThreeDynkin.
In the paper, we will throughout use the collective 10-dimensional
spinor index $\alpha$, which means that
$\gamma^i\gamma^a=-\gamma^a\gamma^i$
(the commuting situation is achieved by letting
$\tilde\gamma^a=\gamma^a\gamma^5$). This simplifies the index
structure and minimises the occurrences of
$\gamma_5=\gamma^0\gamma^1\gamma^2\gamma^3$.

\Figure\DTwoDThreeDynkin{\epsffile{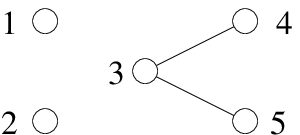}}{Dynkin diagram for
$D_2\times D_3$.}

It is
obvious that the same action, using the same pure spinors, gives a
good description. However, if one looks for supermultiplets that are
inherent to the dimensionally reduced theory, this turns out to be
less natural.

\section\ScalarSuperfieldSection{Gauge-covariant superfields?}Examples
of gauge-covariant multiplets, existing in $D=4$ but not in $D=10$.
are the scalar multiplet, 
starting with the six scalars $\phi_a$ (which of course, on shell,
contains the local degrees of freedom of the SYM multiplet)
and the gauge-invariant stress tensor multiplet, starting
with $\phi_{(a}\phi_{b)'}$ in the traceless symmetric representation
${\bf20}$ of $Spin(6)\approx SU(4)$.

Consider first the multiplet based on the scalars. Since it starts
with a gauge-covariant field, the YM connection will appear only
through its field strength, and the linearised
equations of motion should contain
both $d{\star}F=0$ and $dF=0$.

Using the methods of ref. [\CederwallKarlssonBI], a pure spinor
superfield based on the scalar fields can be formed from the original
one by acting with a ``physical operator'',
$$
\tilde\Phi_a=\hat\phi_a\Psi\komma\eqn
$$
where
$$
\hat\phi_a=-\fr4(\lambda\lb)^{-1}(\lb\gamma_aD)
+\fr{32}(\lambda\lb)^{-2}(\lb\gamma_a{}^{mn}d\lb)N_{mn}\punkt\eqn
$$
($N_{mn}$ is the operator $(\lambda\gamma_{mn}{\*\over\*\lambda})$,
which respects the pure spinor constraint.)
Such a field will obey all the desired relations, thanks to the
identity $\{Q,\hat\phi_a\}=-(\lambda\gamma_a\hat\chi)$, where
$\hat\chi^\alpha$ is the physical operator for the fermion fields (see
ref. [\CederwallKarlssonBI]).
This is however not enough, since it only tells us how to extract
fields from $\Psi$, not how to define them from scratch.

If we inspect the zero-mode cohomology of a field $\tilde\Phi_a$, which is
assumed to enjoy the shift symmetry 
[\CederwallKarlssonBI] $\tilde\Phi_a\approx
\tilde\Phi_a+(\lambda\gamma_a\xi)$, we
obtain a list with some problems. The {\it desired} cohomology is listed in
Table \DFourScalarTable.
In the table, there are the scalar fields, the spinors, and the YM
field strength at ghost number 0. At ghost number $-1$, one should find
the equations of motion for the spinors and scalars, both the YM equation of
motion and the Bianchi identity, and nothing more.
This is however {\it not} what happens.
The field content at ghost number 0 turns out to be correct. All
equations of motion (antifields) are also obtained at ghost number
$-1$. However, there is undesired cohomology, in a large number of
representations, at the
antifield level. The phenomenon starts with the representation $(0)(0)(200)$ at
$\lambda\theta$. Why does this happen? If we had not set $\*_a=0$,
this cohomology could have been reached with a derivative. In order to
get rid of the unwanted cohomology, and reproduce
Table \DFourScalarTable, the pure spinor constraint has to
be modified.

\section\PureSpinorSection{$D=4$, $N=4$ pure spinors}We introduce a
new ``$D=4$, $N=4$ pure spinor'' $\mu^\alpha$, satisfying 
$$
(\mu\gamma^i\mu)=0\komma\eqn
$$
but with no constraint on $(\mu\gamma^a\mu)$. Note that this is the
maximal relaxation of the $D=10$ pure spinor constraint allowed after
dimensional reduction, since, with $q=\mu^\alpha D_\alpha$,
$$
q^2=\fr2\mu^\alpha\mu^\beta\{D_\alpha,D_\beta\}=-(\mu\gamma^i\mu)\*_i\punkt\eqn
$$

The pure spinor constraint $(\mu\gamma^i\mu)=0$ is
irreducible. Remember that the effective number of constraints on a
$D=10$ pure spinor is 5. $D=4$ is indeed the highest dimensionality
where a relaxed constraint (not being equivalent to the $D=10$
constraint) is possible. The number of degrees of freedom in $\mu$ is
12, compared to 11 for $\lambda$.
The partition function
for such a pure spinor is obviously
$$
Z(t)={(1-t^2)^4\over(1-t)^{16}}={(1+t)^4\over(1-t)^{12}}\punkt\eqn
$$
The numerator $(1-t^2)^4=1-4t^2+6t^4-4t^6+t^8$ relates to the
cohomology of a scalar superfield in Section \MultipletsSection.
The partition function (and the corresponding cohomology) is easily
refined to its full representation content; then the numerator
becomes the ghost partition
$$
Z_{\hbox{\sixrm
gh}}=\bigoplus_{p=0}^4t^{2p}\wedge^p(1)(1)(000)\punkt
\Eqn\GhostPartition
$$
The superalgebra associated to the pure spinor partition function, in
the sense of ref. [\CederwallPalmkvistBorcherds], is the superconformal
algebra
${\frak su}(2,2|4)$. It is unclear if this is a coincidence. 

The integration measure also changes. Again, there is a holomorphic
top-form. Instead of behaving like
$\lambda^{-3}d\lambda^{11}$, it will go as
$\mu^{-4}d\mu^{12}$. Its explicit form is related to the
top cohomology of Table \MuPsiTable\ (see section \MultipletsSection),
which has a representative in terms of minimal variables
$\omega\sim\epsilon_{i_1i_2i_3i_4}(\mu\gamma^{i_1}\theta)
(\mu\gamma^{i_2}\theta)(\mu\gamma^{i_3}\theta)(\mu\gamma^{i_4}\theta)$.
The measure is 
$$
\Omega\sim(\mu\mb)^{-4}\epsilon_{i_1i_2i_3i_4}\epsilon_{\alpha_1\ldots\alpha_{16}}
(\gamma^{i_1}\mb)^{\alpha_1}\ldots(\gamma^{i_4}\mb)^{\alpha_1}
d\mu^{\alpha_5}\ldots d\mu^{\alpha_{16}}\punkt\eqn
$$
A measure $[dZ]=d^4x\,d^{16}\theta\,\Omega$ is bosonic and carries
dimension 0 and ghost number $-4$.

As usual, a BRST-invariant regulator [\BerkovitsNonMinimal]
$e^{-\{Q,\xi\}}$ is needed in the
measure (or in the representatives of the cohomologies).
The regulator can
conveniently be chosen as $\xi=\alpha(\mb\theta)$, giving
$\{Q,\xi\}=\alpha((\mu\mb)+(d\mb\theta))$.
The integration will be independent of $\alpha>0$, and letting
$\alpha\rightarrow\infty$ localises the integral at $\mu=0$. The
regulator both makes the bosonic integration finite at
$(\mu\mb)\rightarrow\infty$, and saturates the form degree.
It is then straightforward to verify that the regulated integral
$\int[dZ]\omega$ gives a finite number.

\section\MultipletsSection{Cohomologies and supermultiplets}The
irreducibility of the constraint implies that the cohomology in a 
scalar field, which can be read off from eq. (\GhostPartition),
changes to a ``trivial'' one, given in
Table \MuPsiTable. This means that pure spinor cohomology becomes de
Rahm cohomology, and only flat connections are produced.
This may seem disastrous, but we will soon see how it is remedied in
an action.

\Table\MuPsiTable{$$
\vtop{\eightpoint\baselineskip16pt\lineskip0pt
\ialign{
$\hfill#\quad$&$\,\hfill#\hfill\,$&$\,\hfill#\hfill\,$&$\,\hfill#\hfill\,$
&$\,\hfill#\hfill\,$&$\,\hfill#\hfill$&$\,\hfill#\hfill$\cr
            \hbox{gh\#}=&1    &0    &-1    &-2   &-3  &-4\cr
\hbox{dim}=0&(0)(0)(000)&\phantom{(0)(0)(000)}&\phantom{(0)(0)(000)}
    &\phantom{(0)(0)(000)}&\phantom{(0)(0)(000)}&\phantom{(0)(0)(000)}\cr
        \fr2&\bullet&\bullet&       &       &    \cr
           1&\bullet&(1)(1)(000)&\bullet&       &    \cr
       \Fr32&\bullet&\bullet&\bullet&\bullet&      \cr
           2&\bullet&\bullet
              &\raise5pt\vtop{\baselineskip=9pt\ialign{\hfill$#$\hfill\cr
        (0)(2)(000)\cr(2)(0)(000)\cr}}&\bullet&\bullet\cr
       \Fr52&\bullet&\bullet&\bullet&\bullet&\bullet&\bullet\cr
           3&\bullet&\bullet&\bullet&(1)(1)(000)&\bullet&\bullet\cr
       \Fr72&\bullet&\bullet&\bullet&\bullet&\bullet&\bullet\cr
           4&\bullet&\bullet&\bullet&\bullet&(0)(0)(000)&\bullet\cr
       \Fr92&\bullet&\bullet&\bullet&\bullet&\bullet&\bullet\cr
}}
$$}{The $\mu$ zero-mode cohomology of a 
$D=4$, $N=4$ scalar superfield $\Psi$.}

Concerning the gauge-covariant field $\Phi_a$ based on the physical
scalars, it is straightforward to verify that
the modified zero-mode cohomology does not contain the unwanted
representations. The zero-mode cohomology is precisely the one
given in Table \DFourScalarTable. The shift
symmetry used is $\Phi_a\approx\Phi_a+(\mu\gamma_a\xi)$.

\Table\DFourScalarTable{$$
\vtop{\eightpoint\baselineskip16pt\lineskip0pt
\ialign{
$\hfill#\quad$&$\,\hfill#\hfill\,$&$\,\hfill#\hfill\,$&$\,\hfill#\hfill\,$
&$\,\hfill#\hfill\,$&$\,\hfill#\hfill$\cr
            \hbox{gh\#}=&0    &-1    &-2    &-3  \cr
\hbox{dim}=1&(0)(0)(100)&\phantom{(00000)}&\phantom{(00000)}
&\phantom{(00000)}\cr
        \Fr32&(0)(1)(010)\,\,(1)(0)(001)&\bullet&       &       &    \cr
           2&(0)(2)(000)\,\,(2)(0)(000)&\bullet&\bullet&   &    \cr
       \Fr52&\bullet
       &(0)(1)(001)\,\,(1)(0)(010)
       &\bullet&\bullet&      \cr
           3&\bullet&(0)(0)(100)\,\,2(1)(1)(000)&\bullet&\bullet\cr
       \Fr72&\bullet&\bullet&\bullet&\bullet\cr
           4&\bullet&\bullet&2(0)(0)000)&\bullet\cr
       \Fr92&\bullet&\bullet&\bullet&\bullet\cr
}}
$$}{The $\mu$ zero-mode cohomology of the
$D=4$, $N=4$ scalar superfield.}

The same procedure can be performed for the stress tensor multiplet. A
symmetric traceless pure spinor superfield $S_{ab}(\theta,\mu)$
has the cohomology
given in Table \StressTensorTable. Using the $D=10$ pure spinor $\lambda$
again gives extra unwanted cohomology.
The shift symmetry $S_{ab}\approx S_{ab}+(\mu\gamma_{(a}\xi_{b)})$
with $\xi_a^\alpha$ an irreducible vector-spinor, has been used.
In addition to the fields in the stress tensor multiplet, their
differential constraints, corresponding to the appropriate
conservation laws
(including the R-symmetry current) are correctly reproduced.
The components of the stress tensor multiplet are described in detail in
Appendix \StressTensorAppendix. 

\Table\StressTensorTable{$$
\vtop{\eightpoint\baselineskip28pt\lineskip0pt
\ialign{
$\hfill#\quad$&$\,\hfill#\hfill\,$&$\,\hfill#\hfill\,$&$\,\hfill#\hfill\,$
&$\,\hfill#\hfill\,$&$\,\hfill#\hfill$\cr
            \hbox{gh\#}=&0    &-1    &-2     \cr
\hbox{dim}=2&(0)(0)(200)&\phantom{(0)(0)(000)}&\phantom{(0)(0)(000)}
                               &\phantom{(0)(0)(000)}\cr
        \Fr52&
        (0)(1)(110)\,\,(1)(0)(101)&\bullet\cr
      3&\raise11pt\vtop{\baselineskip=10pt\ialign{\hfill$#$\hfill\cr
           (0)(0)(002)\,\,(0)(0)(020)\cr
           (0)(2)(100)\,\,(2)(0)(100)\cr
           (1)(1)(011)\cr}}&\bullet&\bullet       &       &    \cr
       \Fr72&\raise5pt\vtop{\baselineskip=10pt\ialign{\hfill$#$\hfill\cr
        (0)(1)(010)\,\,(1)(0)(001)\cr
           (1)(2)(001)\,\,(2)(1)(010)\cr}}
                       &\bullet& \bullet      &       &    \cr
           4&2(0)(0)(000)\,\,(2)(2)(000)&(0)(0)(011)&\bullet\cr
       \Fr92&\bullet&(0)(1)(001)\,\,(1)(0)(010)&\bullet\cr
           5&\bullet&(1)(1)(000)&\bullet\cr
       \Fr{11}2&\bullet&\bullet&\bullet\cr
}}
$$}{The $\mu$ zero-mode cohomology of the 
$D=4$, $N=4$ stress tensor superfield.}

\section\ActionSection{A new action for $D=4$, $N=4$ SYM}Can an action
be written in the new pure spinor space? 
The BV
Lagrangian density should have dimension 0 and ghost number 4.
A Chern--Simons-like action like eq. (\DTenAction) is no longer
possible. Neither would it be meaningful, considering that $\Psi$ no
longer carries the cohomology of the SYM multiplet. The local
degrees of freedom reside entirely in $\Phi_a$.
The shift symmetry $\delta\Phi_a=(\mu\gamma_a\xi)$ should be implied
by the action. This can only happen if $\Phi_a$ appears in the
combination
$(\mu\gamma^a\mu)\Phi_a$. Then the shift symmetry is present thanks to
the pure spinor constraint and the usual $D=10$ Fierz identity:
$$
(\gamma_a\mu)_\alpha(\mu\gamma^a\mu)=(\gamma_m\mu)_\alpha(\mu\gamma^m\mu)=0
\punkt\eqn
$$
This combination has dimension 0 and ghost number 2. A ``kinetic
term'', implying the linearised equations of motion for $\Phi_a$ and
$\Psi$, is $(\mu\gamma^a\mu)\Tr\Phi_aQ\Psi$.
It can be ``covariantised'' with respect to the connection in $\Psi$
to $(\mu\gamma^a\mu)\Tr\Phi_a(Q\Psi+\Psi^2)$. However, such an action
will still
lead to a flat connection in $\Psi$ through the equation
$Q\Psi+\Psi^2=0$. What is needed is an equation of
motion implying that $\Psi$ contains a connection whose field strength
is the 2-form in the cohomology of $\Phi_a$. This is achieved by the
only remaining possible term,
$(\mu\gamma^a\mu)(\mu\gamma^b\mu)\Tr\Phi_a\Phi_b$, as we will
demonstrate shortly.
The full action is
$$
S=\int[dZ]\Tr\left[(\mu\gamma^a\mu)\Phi_a(Q\Psi+\Psi^2)
+\fr2(\mu\gamma^a\mu)(\mu\gamma^b\mu)\Phi_a\Phi_b\right]
\punkt\eqn
$$
Note that the $\Phi^2$ term contains the traceless symmetric tensor, \ie,
precisely the stress tensor superfield. 

The equations of motion following from the action are
$$
\eqalign{
&Q\Psi+\Psi^2+(\mu\gamma^a\mu)\Phi_a=0\komma\cr
&Q\Phi_a+[\Psi,\Phi_a]=0\punkt\cr
}
\Eqn\EquationsOfMotion
$$

We will now check that the stress tensor term in the action indeed
implies that the field $F_{ij}$ in the cohomology of $\Phi$ becomes
identified with the field strength of the connection in $\Psi$.
Let $\Psi=(\mu\gamma^i\theta)A_i(x)$. Then,
$$
Q\Psi+\Psi^2=\fr2(\mu\gamma^i\theta)(\mu\gamma^j\theta)F^{(A)}_{ij}\komma\eqn
$$
where $F^{(A)}_{ij}=2(\*_{[i}A_{j]}+A_{[i}A_{j]})$.
This should be compared to the field $F_{ij}$, which in the $\Phi_a$
cohomology sits as
$\fr4(\theta\gamma_a\gamma^{ij}\theta)F_{ij}$. We contract this
with $(\mu\gamma^a\mu)$ to obtain
$(\mu\gamma^a\mu)\Phi_a=\fr4(\mu\gamma^a\mu)
(\theta\gamma_a\gamma^{ij}\theta)F_{ij}$.
This is cohomologically equivalent to
$\fr2(\mu\gamma^a\theta)(\mu\gamma_a\gamma^{ij}\theta)$. On the other
hand, we can use a 10-dimensional Fierz identity together with the
pure spinor constraint to obtain
$$
(\mu\gamma^a\theta)(\mu\gamma_a\gamma^{ij}\theta)
+(\mu\gamma^k\theta)(\mu\gamma_k\gamma^{ij}\theta)=-\fr2(\mu\gamma^a\mu)
(\theta\gamma_a\gamma^{ij}≥\theta)\punkt\eqn
$$
Taken together, 
$$
(\mu\gamma^a\mu)\Phi_a\approx-\fr4(\mu\gamma^k\theta)
(\mu\gamma_k\gamma^{ij}\theta)F_{ij}
=-\fr2(\mu\gamma^i\theta)(\mu\gamma^j\theta)F_{ij}
-\fr4(\mu\gamma^k\theta)(\mu\gamma_k{}^{ij}\theta)F_{ij}
\punkt\eqn
$$
The second term is trivial.

The first equation in (\EquationsOfMotion) thus sets $F^{(A)}=F$. This
also implies that the second equation of motion gives covariant
derivatives with the right connection, and the on-shell equivalence
with $D=4$, $N=4$ SYM theory is established.


\vfill\eject

\section\ConclusionsSection{Conclusions}Modifying the pure spinor
constraint to $(\mu\gamma^i\mu)=0$, $i=0,\ldots,3$, turns out to be
fruitful in $D=4$, $N=4$ SYM. It is shown that it is necessary for a
correct description of gauge-covariant multiplets, \ie, any
superfields based on the scalar fields. Even if the ghost number 0
sector is the right one when 10-dimensional pure spinors are used, the
presence of ``false'' cohomology at lower ghost number disqualifies
such fields for quantum calculations.

The new action given has all propagating degrees of freedom located in
the superfields $\Phi_a$ based on the scalar fields, although the
gauge connection is shared with the gauge superfield $\Psi$.
The action shares the property with all other pure spinor superfield
actions that it is of lower degree in fields than the corresponding
component field action (even in cases with non-polynomial component
action, the pure spinor superfield action becomes polynomial
[\PureSGII,\CederwallKarlssonBI,\ChangDeformationsII]).
In the present case, the only cubic terms come
from a field strength and a minimal coupling.

Deriving Feynman diagrams and computing amplitudes will involve gauge
fixing. Since the fields use have a natural interpretation in the BV
setting, all fields will need gauge fixing. In the superfield
$\Phi_a$, which has no ghost zero-mode cohomology, this will amount to
eliminating antifields. This should be done using a $b$ operator
[\BerkovitsNonMinimal],
which will be very similar to the one in the $D=10$ formalism.

The action presented here then looks like a good starting point for
calculation of correlators of \eg\ the stress tensor multiplet and the
Konishi multiplet, or of more complicated operators.

\appendix\StressTensorAppendix{The scalar and stress tensor
multiplets}The ghost number 0 part of the field $\Phi_a$ in minimal
pure spinor variables is a standard
superfield $\phi_a(x,\theta)$. The condition $Q\Phi_a=0$ gives
$$
\eqalign{
D_\alpha\phi_a&=-(\gamma_a\chi)_\alpha\komma\cr
D_\alpha\chi^\beta&=\fr2(\gamma^{ij})_\alpha{}^\beta F_{ij}
       +(\gamma^{ia})_\alpha{}^\beta\*_i\phi_a\komma\cr
D_\alpha F_{ij}&=2(\gamma_{[i}\*_{j]}\chi)_\alpha\komma\cr
}\Eqn\phiEquations
$$
where
$$
\eqalign{
\square\phi_a&=0\komma\cr
(\dslash\chi)_\alpha&=0\komma\cr
\*^jF_{ij}&=0\komma\cr
\*_{[i}F_{jk]}&=0\punkt\cr
}
$$

The stress
tensor multiplet, in
a superfield $S_{ab}$, is obtained as the square of the scalar
superfield,
$$
S_{ab}=\Tr(\phi_a\phi_b-\fr6\delta_{ab}\phi^c\phi_c)\punkt\eqn
$$
Acting with consecutive fermionic covariant derivatives and using the
equations (\phiEquations) gives the
linearised multiplet with the field content of
Table \StressTensorTable.
Concretely, suppressing the traces,
$$
\eqalign{
D_\alpha S_{ab}&=2(\gamma_{[a}Y_{b]})_\alpha\komma\cr
D_\alpha Y_a^\beta&=-\fr6(\gamma_a\gamma^{bcd})_\alpha{}^\beta J_{bcd}
   -(\gamma^i(\gamma^b\delta^c_a-\fr6\gamma^{bc}\gamma_a))_\alpha{}^\beta
   J_{ibc}\cr
&\quad   +\fr2(\gamma^{ij}(\delta_a^b-\fr6(\gamma^b\gamma_a))_\alpha{}^\beta J_{ijb}
        +\ldots\komma\cr
D_\alpha J_{abc}&=\fr8(\gamma_{abc}K)_\alpha+\ldots\komma\cr
D_\alpha J_{iab}&=-\fr2(\gamma_{ab}K_i)_\alpha+\ldots\komma\cr
D_\alpha
J_{ija}&=\fr4(\gamma_{aij}K)_\alpha+(\gamma_{a[i}K_{j]})_\alpha+\ldots\komma\cr
D_\alpha K^\beta&=2\delta_\alpha{}^\beta L-2(\gamma_5)_\alpha{}^\beta
I+\ldots\komma\cr
D_\alpha K_{i\beta}&=2\gamma^j_{\alpha\beta}T_{ij}+\ldots\komma\cr
}\eqn
$$
where the ellipses denote terms with bosonic derivatives on
lower-dimensional components, arising from symmetrised fermionic
derivatives, and where
$$
\eqalign{
Y_a^\alpha&=\phi_a\chi^\alpha-\fr6\phi_b(\gamma_a\gamma^b\chi)^\alpha\komma\cr
J_{abc}&=\fr{16}(\chi\gamma_{abc}\chi)\komma\cr
J_{iab}&=\phi_{[a}\*_{|i|}\phi_{b]}+\fr8(\chi\gamma_{iab}\chi)\komma\cr
J_{ija}&=\phi_aF_{ij}-\fr8(\chi\gamma_{ija}\chi)\komma\cr
K^\alpha&=\fr2(\gamma^{ij}\chi)^\alpha F_{ij}\komma\cr
K_{i\alpha}&=(\gamma^j\chi)_\alpha F_{ij}
    -\fr4(\gamma_i\gamma^{jk}\chi)_\alpha F_{jk}\cr
 &\quad   -\fr3(\gamma^a\*_i\chi)_\alpha\phi_a
    +\Fr23\left[(\gamma^a\chi)_\alpha\*_i\phi_a
                -\fr4(\gamma_i\gamma^{ja}\chi)_\alpha\*_j\phi_a\right]\komma\cr
L&=\fr4F^{ij}F_{ij}\komma\cr
I&=\fr8\epsilon^{ijkl}F_{ij}F_{kl}\komma\cr
T_{ij}&=\fr2(F_i{}^kF_{jk}-\fr4\eta_{ij}F^{kl}F_{kl})-\fr4(\chi\gamma_{(i}\*_{j)}\chi)
\cr
&\quad+\fr3(\*_i\phi^a\*_j\phi_a-\fr4\eta_{ij}\*^k\phi^a\*_k\phi_a)
    -\fr6\phi^a\*_i\*_j\phi_a\komma\cr
}\eqn
$$
subject to the equations
$$
\eqalign{
\*^iJ_{iab}&=0\komma\cr
\*^iK_{i\alpha}&=0\komma\cr
\*^jT_{ij}&=0\punkt\cr
}\eqn
$$

\acknowledgements
I would like to thank Evgeny Ivanov for discussions on superspace
formulations of $D=4$, $N=4$ SYM, and Emery Sokatchev for asking a
question which initiated the present investigation.

\refout

\end